\documentclass[12pt]{article}

\usepackage{graphicx}
\usepackage{pslatex}
\usepackage{float}
\usepackage{color}
\usepackage[table]{xcolor}
\usepackage{array}
\usepackage{amssymb}
\usepackage{amsmath}
\usepackage{euscript}

\usepackage{pxfonts}
\usepackage{ifthen}
\usepackage{hyperref}

\usepackage{longtable}
\usepackage{eso-pic}

\usepackage{xspace}
\usepackage[paperwidth=21.0cm,paperheight=29.7cm,textwidth=15cm,textheight=20cm,top=2cm,left=2cm,right=2cm,bottom=2cm]{geometry}

\usepackage{amssymb,amsmath,subfig,color,mathrsfs}
\usepackage{caption}
\usepackage{subfig}

\newcommand{\be}{\begin{equation}} \newcommand{\ee}{\end{equation}}
\renewcommand{\ee}{\end{equation}}
\newcommand{\bes}{\begin{equation*}}
\newcommand{\ees}{\end{equation*}}
\renewcommand{\vec}[1]{\boldsymbol #1} 
\newcommand{\pext}{p_{\mbox{\tiny ext}}}

\usepackage{soul,ulem,color,xspace,bm}
\renewcommand{\vec}[1]{\boldsymbol{#1}}

\definecolor{darkgreen}{rgb}{0,0.5,0}

\renewcommand{\vec}[1]{\boldsymbol{#1}}

\definecolor{refcolor}{rgb}{0.3,0.3,0.7}
\hypersetup{
    pdftitle={Fluid flow across a wavy channel brought in contact},
    pdfauthor={Andrei G. Shvarts, Vladislav A. Yastrebov},
    pdfsubject={Fluid flow in contact interface},
    pdfcreator={LaTeX, Kile},
    pdfkeywords={wavy channel, contact, fluid flow, sealing, fluid-solid coupling, monolithic approach, finite element analysis},
    pdfnewwindow=true,
    colorlinks=true,
    linkcolor=refcolor,          
    citecolor=refcolor,          
    filecolor=refcolor,      
    urlcolor=refcolor           
}
\usepackage{breakcites}

\title{Fluid flow across a wavy channel brought in contact}
\author{ Andrei G. Shvarts, Vladislav A. Yastrebov}
\date{\small \textit{MINES ParisTech, PSL Research University, Centre des Mat\'eriaux, CNRS UMR 7633, BP 87, 91003 Evry, France}}

\begin{document}
\maketitle

\begin{flushleft}
 \large{\bf Abstract.}\normalsize
\end{flushleft}

\noindent 
A pressure driven flow in contact interface between elastic solids with wavy surfaces is studied. We consider a strong coupling between the solid and the fluid problems, which is relevant when the fluid pressure is comparable with the contact pressure. An approximate analytical solution is obtained for this coupled problem.
A finite-element monolithically coupled framework is used to solve the problem numerically. A good agreement is obtained between the two solutions within the region of the validity of the analytical one. A power-law interface transmissivity decay is observed near the percolation. Finally, we showed that the external pressure needed to seal the channel is an affine function of the inlet pressure and does not depend on the outlet pressure.

\begin{flushleft}
\end{flushleft}

\section{Introduction} 

The problem of a thin fluid flow in narrow interfaces between contacting or slightly separated surfaces occurs in different applications. The first example is the sealing problem: seals are used to minimize or prevent leakage of fluids from and into internal chambers of numerous engineering systems, such as gas cylinders, water circuits, lubricated bearings and gears, heat engines and others. Dynamic and static seals are distinguished, the former seal interfaces between surfaces with no relative motion, the latter deal with relatively moving surfaces. Contact and non-contact seals are also distinguished: the former possess contacting parts in the sealing interface, the latter do not.

Hydraulic fracturing is another application which involves interaction of fluid and solid with possible contact between crack faces or/and with a third body, like sand particles~\cite{bazant2014jam}. The fluid extraction of shale gas and oil from rocks represents an antipodal problematic to sealing applications, but the physics and fluid-solid coupling remain identical. A slightly different problem involving fluid, solid and contact appears in fatigue-crack growth in lubricated rolling or cyclically sliding contacts~\cite{Bower_1988}. Such an interaction between fluids and solids in contact can be also found in poromechanics~\cite{Coussy_2004} and at larger scales in landslides~\cite{viesca2012jgr}, slip in pressurized faults~\cite{garagash2012jgr}, bazal sliding of glaciers~\cite{fischer1997aog}, and in other applications.

Depending on the application, the fluid can be considered compressible or incompressible (both in gas and liquid states), and it might flow under capillary effect or pressure difference. Fluids in liquid state at high pressure may evaporate due to a pressure decrease along the fluid path or due to temperature increase induced by frictional heating, which results in a mixed gas-liquid-solid problem: a notable example is cavitation in lubrication problems. Compressibility and viscosity of fluid can depend significantly on pressure and temperature within a certain range of loading parameters.
Polymers are used in most sealing applications, however, due to fluid-uptake, chemical and thermal degradation, and also the glass transition, their usage is limited; several applications require usage of metal-to-metal contacts in seals. In nature, the problems of interfacial fluids in contact interfaces are relevant for rock materials in terms of hydrogeology, shale gas and oil extraction as well as fracking, and also magma rise in volcanology~\cite{lister1991fluid}. In biological systems, the relevant materials are soft tissue and the applications include circulation of blood and other fluids in organisms.

Inevitable roughness, sometimes complemented by on-purpose patterning of engineering and natural surfaces, affects their sealing properties. Inversely, the presence of a fluid in the contact interface may affect the mechanical properties of seals by adding extra load-carrying capacity, changing the interface stiffness and the friction coefficient. For soft materials and high fluid pressures, interface fluids can considerably deform the solids in elasto-hydrodynamic lubrication, however it may be also relevant for static seals~\cite{muller1998b}. In {\it contact} seals this nonlinear fluid-solid interaction is intensified by nonlinear contact constraints. This coupling presents the topic of the current study.

The roughness of contacting surfaces~\cite{whitehouse2010handbook,thomas1999b} has strong implications in mechanics and physics of contact: adhesion, friction and wear in dry and lubricated contacts are controlled to a great extent by parameters of the roughness of contacting solids. 
Mass and energy transport through and across contact interfaces strongly depend on the surface roughness, for instance: electric contact resistance, heat conduction between contacting solids and the sealing problem -- the topic of the interest of the present paper. 
The roughness, or more generally the surface geometry, may contain some deterministic features (turned surfaces, patterned surfaces~\cite{prodanov2013tl}) or be purely random, self-affine~\cite{nayak1971tasme} down to atomistic scale~\cite{krim1995ijmpb,ponson2006ijf}. Surface morphology may be determined by surface processing (polishing, work-hardening), underlying microstructure and its deformation (grain boundaries, plasticity induced roughness~\cite{siska2006finite}, persistent slip marks~\cite{bao1989am}, rumpling~\cite{tolpygo2000surface}), corrosion and oxidation; for coatings the roughness is determined by the deposition method (physical vapor deposition, gas dynamic cold spray deposition, electroplating and others), in biology the surface is determined by the tissue growth processes and related instabilities or assigned functionalities~\cite{BENAMAR2010935,LI2011758,wang2015three}. 
The resulting surface morphology may be rather complex and span over many scales from atomistic to structural ones, it can be characterized by numerous parameters, such as standard deviation of heights and height gradient, height distribution, in particular its kurtosis and skewness, power spectral density, spectral moments, etc. 
For mechanical contact problems, since in most applications only the highest asperities come in contact, an approximation of the roughness by a number of isolated spherical or elliptic asperities results in a rather accurate and helpful model~\cite{archard1957,greenwood1966,bush1975elastic}. On the other hand, the fluid flow through the free volume\footnote{By free volume here we mean the separation field between contacting surfaces.} is mainly affected by ``deeper'' surface features: grooves, valleys and dimples.

Numerical analysis of the fluid flow through contact interfaces was carried out between real~\cite{amyot2007tpm,vallet2009ti_a,vallet2009ti,durand2012phd} or model rough topographies~\cite{plouraboue2004pof,dapp2012prl,paggi2015w}. 
In contrast to the complexity and lack of scale separation of nominally flat realistic surface roughness, surface patterning allows to use the concept of scale separation to a certain extent and to limit the analysis to major geometrical features of the surface~\cite{sahlin2010pime_a,sahlin2010pime_b,rafols2016modelling}. 
On the other hand, analysis of simple models of surface geometry, for example wavy and bi-wavy models~\cite{westergaard1939jam,kuznetsov1985w,johnson1985ijms,johnson1995adhesion,carbone2004jmps,yastrebov2014tl,dapp2015contact,xu2015jt,Matsuda2014ti}, helps to understand better local deformation mechanisms in rough contact and the role of patterning for macroscopic behavior. 
A wavy channel also serves as an important test model in fluid flow analysis~\cite{nishimura1984flow,brown1995grl,sui2010ijhmt,nivceno2001ijhff,dapp2015contact}. 
Here we also consider a periodic wavy channel, but contrary to other flow studies, it is brought in mechanical contact with a rigid flat and the fluid flows across the wavy section in channels delimited by mechanical contact zones. In the first approximation this model represents a ``rough'' surface with parallel grooves.

To analyze the fluid flow in the contact interface, two problems should be solved: a mechanical problem of the contact between solids with rough surfaces and the problem of the fluid flow through the resulting free volume. If during the loading process the fluid pressure is negligible in comparison to the contact pressure, it is possible to assume that both physics are weakly coupled. This implies that they can be solved separately: firstly, the solid contact problem is solved to obtain the deformed geometry of the channels. This geometry is then passed to the fluid solver, which resolves the fluid flow under the assumption of rigid boundaries. This solution strategy leads to a one-way coupling, whereby the contact problem is independent of the fluid pressure, while the fluid problem depends on the geometry computed by the solid solver.
However, it is not rare that in the load interval of interest a stronger coupling between fluid and solid equations is required. For example, it is the case when the local contact pressure is comparable with the hydrostatic fluid pressure. Note that it is always the case near edges of contact zones at which the contact pressure (in the uncoupled case) decreases to zero as $\sim\sqrt{\xi}$, when the distance from the contact edge $\xi$ decreases $\xi\to0$~\cite{manners2006w}. The so-called two-way coupling strategy allows to take into account the fluid pressure distribution and its effect on the deformation of the solid and vice versa: the effect of the elastic deformation on the fluid flow. 

In many industrial applications the thickness of the free volume interface between contacting surfaces is quite small, and the flow is often laminar even for gas~\cite{muller1998b}. 
Nevertheless, considering turbulent flow could be essential for more demanding applications, like drag delivery, microfluidic chemical reactors, etc.
Moreover, we postulate that the variation of the mechanical loading conditions is slow enough compared to characteristic time of the fluid flow, so the flow is assumed to be stationary;
capillary actions are neglected, only pressure driven flow is considered. This two-dimensional flow through the contact interface is accurately described by the Reynolds equation~\cite[e.g.]{muller1998b,Hamrock_2004}, which is used in this study.

The paper is structured as follows:  in Section~\ref{sec:setup} we formulate the problem to be solved; in Section~\ref{sec:west} we recall classical solutions for a wavy profile with a pressurized fluid present in the interface; in Section~\ref{sec:approximate} we obtain an approximate analytical solution; in Section~\ref{sec:coupling} a monolithic numerical scheme which couples the fluid and solid equations is briefly outlined; finally, in Section~\ref{sec:results} the numerical results are presented and discussed, and in Section~\ref{sec:conclusion} we make conclusions.


\begin{figure}
 \begin{center}
  \includegraphics[width=1\textwidth]{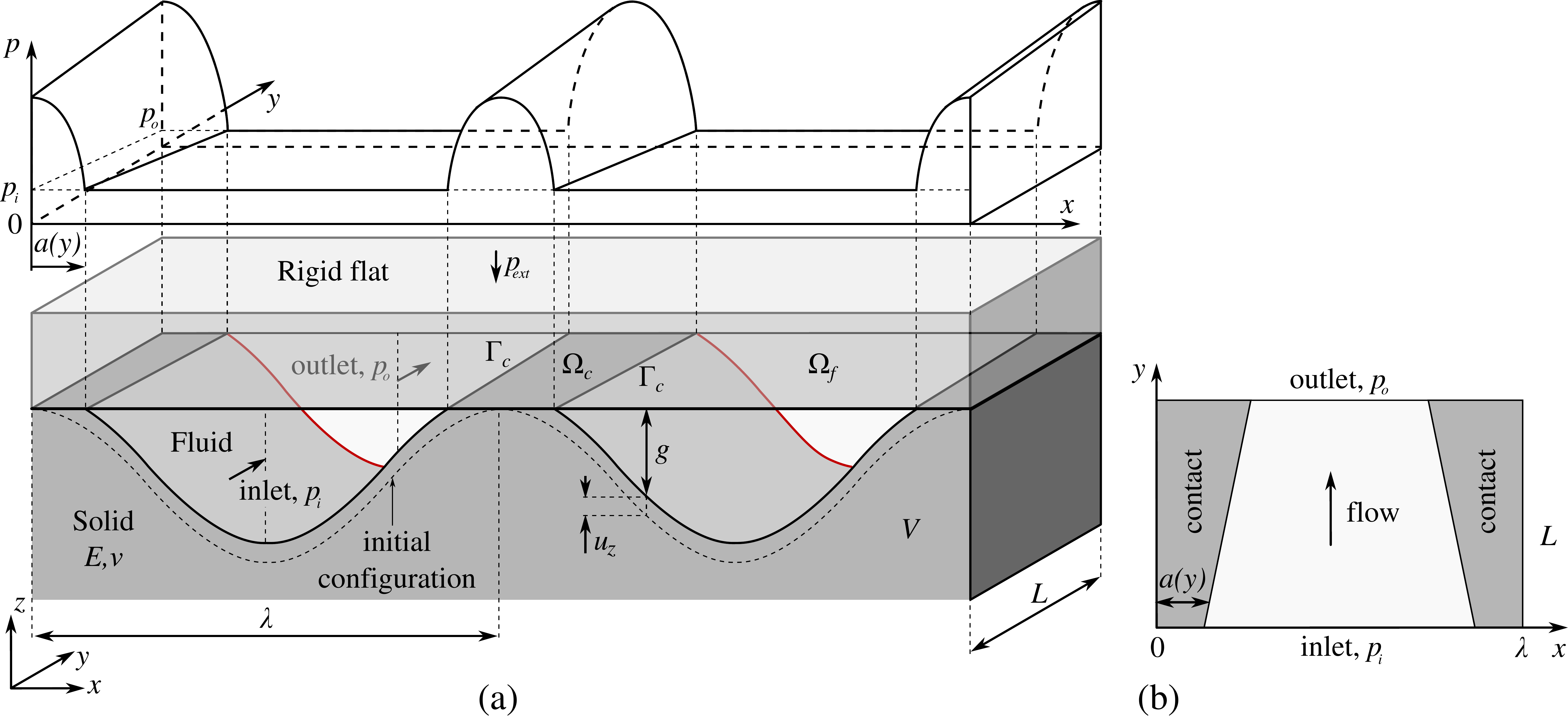}
 \caption{\label{fig:2} (a) Problem set-up: an elastic wavy surface comes in contact with a rigid flat, and an incompressible fluid flows under pressure difference from the inlet to the outlet; the resulting interface pressure distribution $p(x,y)$ is sketched. (b) Sketch of the contact interface, note that due to fluid pressure acting on the surface of the solid, the contact patches in the sections parallel to OX are wider at the outlet, than at the inlet, which is associated with the direction of the fluid pressure drop.}
 \end{center}
\end{figure}

\section{Problem set-up\label{sec:setup}}

We consider an array of wavy channels of length $L$ along $OY$ direction (see Fig.~\ref{fig:2}(a)) with a sine-shape section
\be
z(x')=\Delta[1 - \cos{(2x')}],
\label{eq:geom}
\ee
where $x'=\pi x/\lambda$, brought in contact with a rigid flat\footnote{Note that all the discussions are valid not only for an elastic solid with a wavy surface in contact with a rigid flat, but for two elastic solids with the effective wavy roughness given by $z=z_1-z_2+c$, where $z_1,z_2$ determine surface geometries of the two contacting solids, and $c$ is an arbitrary constant. However, for simplicity hereinafter we will assume that an elastic wavy surface is brought in contact against a rigid flat.}, the pressure driven flow in this channel ($\Delta p_f = p_i - p_o$, where $p_i$ and $p_o$ are the inlet and outlet pressures, respectively) of incompressible viscous fluid is governed by the stationary Reynolds equation for the Poiseuille flow.

We assume an isothermal fluid flow at a temperature at which it does not evaporate under the pressure drop on its way from the inlet to the outlet. 
The system of equations to be solved takes the following form:
\begin{align}
   &\nabla\cdot\left[g^3 \nabla p_f\right] = 0 \quad\mbox{ in }\quad\Omega_f \label{eq:setup1}\\
   &\left.p_f\right|_{y=0} = p_i,\quad \left.p_f\right|_{y=L} = p_o,\quad \left.[\nabla p_f \cdot \vec e_x]\right|_{x=0,\lambda/2} = 0	\label{eq:setup2}\\
   &\nabla\cdot\vec\sigma = 0\quad\mbox{ in } \quad V	\label{eq:setup3}\\
   &\left.u_x\right|_{x=0,\lambda} = 0,\quad \left.u_y\right|_{y=0,L} = 0,\quad \left.\sigma_{zz}\right|_{z=-\infty} = -p_{\mbox{\tiny ext}} \label{eq:setup4}\\ 
   &g\ge 0,\quad p-p_f\ge 0,\quad (p-p_f)g = 0\quad\mbox{ in }\quad\partial V,	\label{eq:setup5}
\end{align}
where Eq.~\eqref{eq:setup1} is the Reynolds equation for pressure driven stationary incompressible viscous Poisseuille flow, the distance between immobile walls is given by at least $\mathrm C^1$-smooth gap distribution $g=g(x,y)$ in the domain $\Omega_f$, which is the closure of the solid volume $\partial V$ projected on the rigid flat, and $p_f$ denotes the fluid pressure. 
Eq.~\eqref{eq:setup2} summarizes boundary conditions for the fluid problem: the fixed inlet $p_i$ and outlet fluid pressure $p_o$ and zero flux at crests of the surface resulting from the problem symmetry. Expression $\nabla p_f \cdot \vec e_x$ is a quantity proportional to the fluid flux in the $OX$ direction (orthogonal to the main flow direction), and $\vec e_x$ is a non-zero in-plane vector collinear with the axis $OX$. The fluid flux is given by $\vec q =  -g^3(\nabla\cdot p_f)/12\mu$, where $\mu$ is the dynamic viscosity. 
Prescribing the fluid flux at the inlet/outlet of the fluid domain would only slightly change the numerical treatment, and since for contact static seals the fluid flow with prescribed hydrostatic inlet and outlet fluid pressures presents a more common situation, for the rest of the paper we will stick to this particular boundary condition. 
Eq.~\eqref{eq:setup3} is the momentum balance equation for the quasi-static solid mechanical problem in absence of volumetric forces, while~\eqref{eq:setup4} summarizes boundary conditions for the solid problem, where $\pext$ is the squeezing pressure applied at infinity. Due to the symmetry, horizontal displacements are zero at lateral walls orthogonal to $OX$, which corresponds to the infinite periodic set-up. Vertical walls on the inlet and outlet sides are assumed to remain flat. Conditions~\eqref{eq:setup5} will be explained in the following section.

Note that the mechanical contact pressure $p$ is defined in contact regions $\Omega_c$, whereas the fluid hydrostatic pressure is a complementary field which is defined in the fluid zone $\Omega_f$, thus for every point $p\cdot p_f=0$ with an exception of the contact-fluid interface (contact edge), $\partial \Omega_f\cap\partial\Omega_c$, at which $p = p_f$. Here, the condition on the contact pressure $p-p_f\ge0$ is valid only on the contact edge shared with the fluid boundary $\partial \Omega_f\cap\partial\Omega_c$, as $p_f=0$ in the interior of the contact zone $\Omega_c\setminus\partial \Omega_c$. Note that such formulation
makes possible existence of contact zones with contact pressure smaller than the pressure of surrounding fluid if and only if these zones are separated from the contact-fluid interface by zones with contact pressure higher than the fluid pressure; a suction cup could be given as an example here.
Linear isotropic elasticity is considered for the elastic half-space $V$, so the stress-strain relation is given by the Hooke's law. Finally, we have one unknown vector field in three dimensions, displacements $\vec u(x,y,z)$ in $V$, and one unknown scalar field in two dimensions, which is the hydrostatic fluid pressure $p_f(x,y)$ in $\Omega_f$. 
We assume that full contact is not reached in all sections parallel to $OX$, so the fluid can always circulate.

In Fig.~\ref{fig:2}(a) we also sketch the resulting interface pressure, note that in the contact zones it is not uniform along the $OY$ axis, but rather increasing towards the outlet side, while the width of contact patches in sections orthogonal to $OY$ is also increasing towards the outlet, in accordance to the direction of the fluid pressure drop, see also Fig.~\ref{fig:2}(b).

We study the evolution of the two fields $\vec u$ and $p_f$ with the increasing external pressure $p_{\mbox{\tiny ext}}$. In particular, we are interested to know how the contact profile $a(y)$ delimits the contour of the fluid channel and how its depth $g(x,y)$ changes. In addition, we can determine under which conditions the flow is possible through the interface, i.e. what are the critical values of $p_i,p_o,\pext$ resulting in channel closure.

\section{Wavy profile with pressurized fluid in the interface\label{sec:west}}

\begin{figure}
 \begin{center}
  \includegraphics[width=1\textwidth]{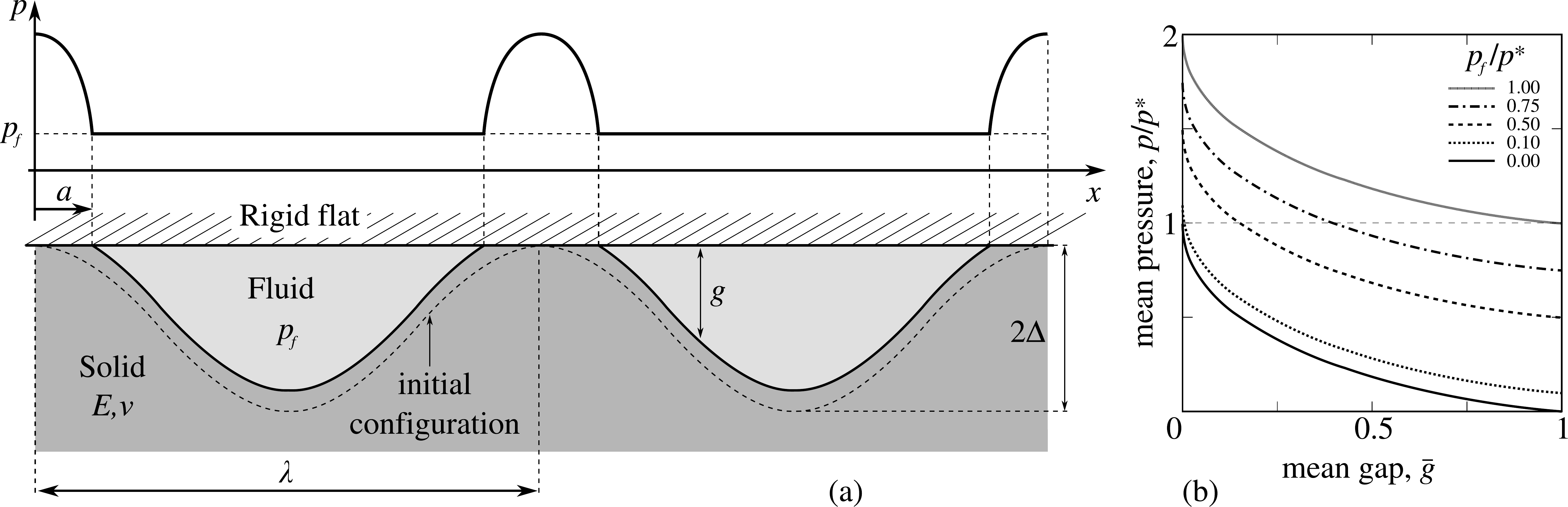}
 \caption{\label{fig:1}(a) - Contact between an elastic wavy profile and a rigid flat with a pressurized fluid in the interface, (b) - variation of the mean pressure on the surface with the normalized gap.}
 \end{center}
\end{figure}

Before making an attempt to solve the three-dimensional problem formulated in the previous section, we focus our attention on a simpler, planar contact problem with a pressurized fluid in the interface. Understanding of this problem will be helpful for the derivation of the approximated solution for the full problem, which is presented in the following section.

An elastic solid with a wavy surface (Fig.~\ref{fig:1}, Eq.~\eqref{eq:geom}) is brought in contact with a rigid flat in a fluid environment, which is retained under a constant pressure $p_f$. A plane strain problem is considered. 
The solid mechanical contact problem was solved for this configuration by Westergaard without fluid pressure ($p_f=0$)~\cite{westergaard1939jam,johnson1985ijms} under the assumption of the infinitesimally small slope of the wavy profile; the pressure distribution in contact region was found to be
\be
 p_{_W}(x',a') = 2 \bar p(a') \frac{\cos(x')}{\sin^2(a')} \sqrt{\sin^2(a')-\sin^2(x')},
\label{eq:west}
\ee
where $x' = \pi x/\lambda, a' = \pi a/\lambda$, and $a$ is the half-width of the contact zone; the mean contact pressure reads as
\be
\bar p(a') = \frac{\pi E^* \Delta}{\lambda} \sin^2(a'),
\label{eq:barp}
\ee 
i.e. $\bar p(a') = \int\limits_0^{a'} p_{_W}(x',a')\,dx'$, and $E^*$ is the effective elastic modulus defined by
$$
\frac{1}{E^*} = \frac{1-\nu_1^2}{E_1}+\frac{1-\nu_2^2}{E_2},
$$
where $E_i,\nu_i$ are the Young's moduli and Poisson's ratios of the two contacting solids $i=1,2$, respectively.
Associated effective displacements (taken with a negative sign) in the contact interface are given by
\be
u_z(x',a') = \begin{cases}
&-\Delta \cos(2x') + C,\mbox{ in contact }\cos(x') > \cos(a')\\
&-\Delta \left[\cos(2x') + 2 \sin(x')h(x',a')\right.-2\sin^2(a')\;\times\\
&\times\;\left.\ln\left(\frac{\sin(x')+h(x',a')}{\sin(a')}\right)\right] + C,\mbox{ out of contact }\cos(x') \le \cos(a'),\\           
          \end{cases} 
\label{eq:west_disp}
\ee
where $h(x',a') = \sqrt{\sin^2(x')-\sin^2(a')}$.

The solution of the contact problem for the same configuration in the pressurized environment was found by Kuznetsov in~\cite{kuznetsov1985w}. If we assume that the fluid pressure acts only vertically\footnote{In~\cite{shvarts2017trap}, this assumption was shown to be too prohibitive for certain applications even if the surface slope is assumed infinitesimal.} and that the profile slope is infinitesimal, the stress state in the contact interface in the presence of the additional fluid pressure, applied outside the contact patches, can be considered as the superposition of the stress state corresponding to the same contact area, but without influence of the fluid, i.e. the Westergaards solution~\eqref{eq:west}, and a uniform field of the fluid pressure $p_f$:
\be
 p(x',a') = p_f + p_{_W}(x',a').
 \label{eq:west_pres}
\ee
A detailed rigorous analysis of the trapped fluid in the contact interface without the assumption of infinitesimal slopes and with the fluid pressure acting normally to the surface was carried out in~\cite{shvarts2017trap}, but here the simplified vision~\eqref{eq:west_pres} is sufficient to investigate the strongly coupled fluid flow along the wavy channel brought in contact.

In the classical Hertzian contact the pressure decreases to zero towards contact edges, but in a pressurized environment such a situation is impossible as the contact would be open by the environmental pressure. So the pressure $p_f$ represents an offset which can be complemented by the mechanical pressure rising in contact. Since a constant $p_f$ does not change the shape of the surface in infinitesimal slope assumption, displacements obtained for pressures~\eqref{eq:west} and ~\eqref{eq:west_pres} differ only by a constant, thus Eq.~\eqref{eq:west_pres} satisfies the contact conditions, which can be formulated for the coupled problem in the following way:
\be
  p-p_f\ge0,\quad g\ge0,\quad (p-p_f)g = 0,
  \label{eq:HSM}
\ee
where $p$ is the contact pressure, $g$ is the gap, and $p_f$ is the pressure of the environment. That is the form which appears in the main system of equations to be solved~\eqref{eq:setup5}.

The relation between the mean gap and the contact force almost does not change compared to the unpressurized case~\cite{johnson1985ijms,kuznetsov1985w}, only a force offset $p_f\lambda$ is added for every period. The mean gap can be computed by integrating $g(x)=z(x)+u_z(x)$ in the non-contact region giving
\be
  \bar g = \Delta\left(1 - \frac{\bar p}{p^*}\left[1-\ln\left(\frac{\bar p}{p^*}\right)\right]\right),\quad\mbox{ for }\bar p \in [0,p^*]
  \label{eq:gap_pres}
\ee
where $p^* = \pi E^* \Delta/\lambda$ is the pressure needed to bring the wavy surface in full contact in the absence of fluid pressure and the mean contact pressure in absence of fluid $\bar p$ is given by Eq.~\eqref{eq:barp}. In Fig.~\ref{fig:1}(b) the full normalized pressure $p/p^* = (\bar p + p_f)/p^*$ is plotted with respect to the mean gap $\bar g$ evolution, where $\bar p+p_f$ is the full pressure applied to the system. It is naturally assumed that the fluid can be squeezed-out from the contact interface.

Based on this planar solution the following preliminary conclusion can be drawn for the three-dimensional problem. If the hydrostatic pressure changes only weakly along the channel, i.e. $p_i \approx p_o$, then the contact lines would remain almost parallel to the axis $OY$ and the derivative of the gap with respect to $y$ may be neglected. 
Then the hydrostatic pressure will be an affine function of the coordinate $y$, i.e. $p_f = p_i + (p_o-p_i)y/L$. The flux, which would have a non-zero component only along $y$ axis, can be readily found as $q_y(x) = -g^3(x)(p_o-p_i)/(12\mu L)$; note that it depends only on the $x$-coordinate.
Thus, naturally for the situation $p_i \approx p_o$ the channel would be sealed at $p_{\mbox{\tiny ext}} \approx p^* + p_o$.

\section{\label{sec:approximate}Approximate solution}

To solve approximately the coupled problem formulated in Eqs.~\eqref{eq:setup1}-\eqref{eq:setup5} we need to make several strong assumptions: we assume (i) that in every section $y=\mbox{const}$, the pressurized Westargaard-Kuznetsov solution~\eqref{eq:west_pres} is satisfied for $a=a(y)$ and $p_f=p_f(y)$. However, it is clear that it should imply that $\partial p_f/\partial x = 0$, which could seem to exclude the channel narrowing. 
But since in the following, the problem will be reduced to a one-dimensional flow along $OY$ axis, this assumption (i) is not contradictory: the fluid pressure can be considered as the mean pressure in the section $$p_f(y) = \frac{1}{\lambda-2a} \int\limits_{a}^{\lambda-a} p_f(x,y)\,dx.$$
We also assume (ii), which is the strongest and the least realistic assumption, that in every section 
\be
\pext = \bar p + p_f = \mbox{const},
\label{eq:pres_cond}
\ee
for that we require that $p_f\le \pext$ in $\Omega$, which is equivalent to require that $p_i\le\pext$.
Another simplification would be (iii) to reduce the two dimensional Reynolds equation to a one-dimensional equation for the average gap~\eqref{eq:gap_pres}, which implies that the hydrostatic pressure is independent of the $x$-coordinate: $p_f=p_f(y)$. Under these assumptions Eq.~\eqref{eq:setup1}-\eqref{eq:setup2} can be rewritten as
\be
  \bar g^3p_f'=C_1,\quad p_f(0)=p_i,\quad p_f(L)=p_o,
  \label{eq:gp2}
\ee
where the prime sign denotes partial derivative with respect to $y$, and $C_1$ is the integration constant. The condition of the zero flow at $x=0,\lambda$ (see Eq.~\eqref{eq:setup1}) is automatically satisfied as $p_f$ is assumed not to depend on $x$. 

\begin{figure}[htb!]
 \includegraphics[width=1\textwidth]{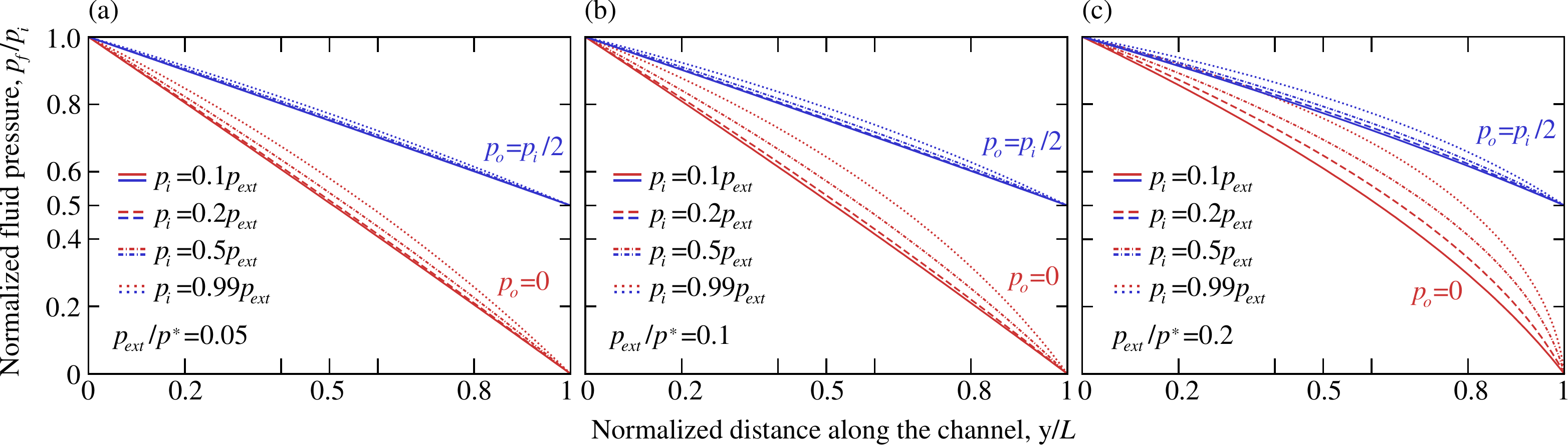}
 \caption{\label{fig:3}Approximate solution~\eqref{eq:py} for the mean hydrostatic fluid pressure distribution along the wavy channel in contact with a rigid flat for $p_o=p_i/2$ and $p_o=0$ (a) $\pext/p^*=0.05$, (b) $\pext/p^*=0.1$, (c) $\pext/p^*=0.2$; the fluid pressure is normalized by the inlet pressure $p_f/p_i$ and the coordinate is normalized by the channel length $y/L$.}
\end{figure}

\begin{figure}[htb!]
  \includegraphics[width=1\textwidth]{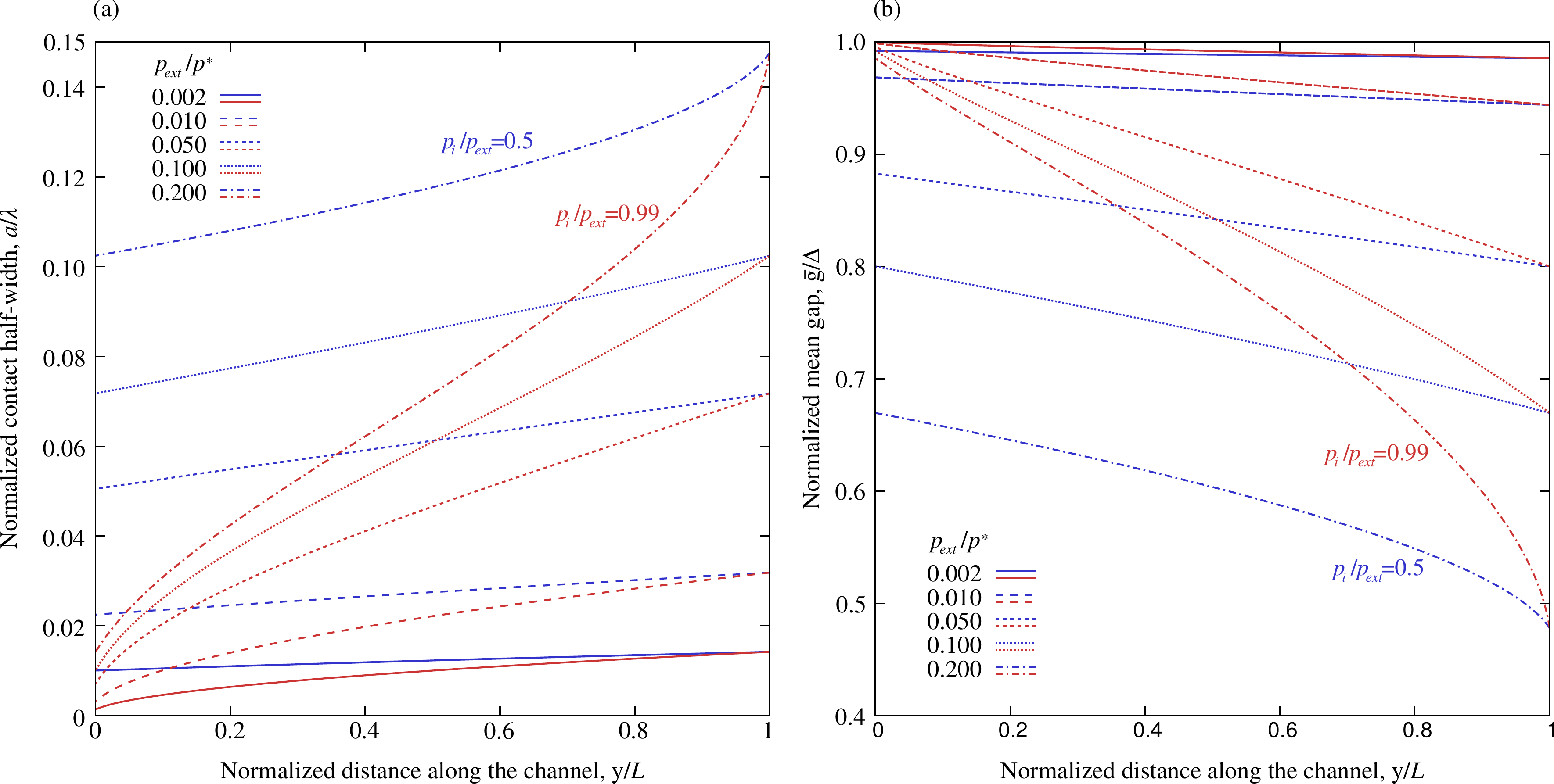}
 \caption{\label{fig:4}Approximate solutions for (a) normalized contact region half-width $a/\lambda$ and (b) normalized mean gap $\bar g/\Delta$ evolution along the channel.}
\end{figure}

Now we can use the relation between the mean gap and the pressure~\eqref{eq:gap_pres} through the relation~\eqref{eq:pres_cond}, which being substituted in~\eqref{eq:gp2} yields:
\be
  \frac{-p^*\Delta^3}{C_1}\int\limits_{\rho_i}^{\rho(y)} \left[1-\rho\left\{1-\ln\rho\right\}\right]^3 d\rho = y,
  \label{eq:int1}
\ee
where $\rho(y) = (\pext-p_f(y))/p^*$, and $\rho_i = \rho(0) = (\pext-p_i)/p^*$. From~\eqref{eq:int1} it follows that $\rho(y)$ should be a monotonically increasing function of $y$ in the range $y\in[0;L]$. The boundary conditions now read as:
$$
  \rho(0) = \rho_i = (\pext-p_i)/p^*,\quad \rho(L) =  \rho_o = (\pext-p_o)/p^*.
$$ 
The integral $I(\rho)=\int \left[1-\rho\left\{1-\ln\rho\right\}\right]^3 d\rho$ from~\eqref{eq:int1} with zero integration constant is evaluated as:
\begin{align}
I(\rho) =& \rho-\alpha_1\rho^2+\alpha_2\rho^3-\alpha_3\rho^4 + \beta_0\rho^2(1-\beta_1\rho+\beta_2\rho^2)\ln(\rho)\nonumber\\
&+ \rho^3(1-\gamma\rho)\ln^2(\rho)+\frac{\rho^4}{4}\ln^3(\rho)\nonumber,
\end{align}
where $\alpha_1=9/4$, $\alpha_2=17/9$, $\alpha_3=71/128$, $\beta_0=3/2$, $\beta_1=16/9$, $\beta_2=13/16$, $\gamma=15/16$. 
The solution cannot be provided in the form $p_f(y)$, but rather $y(p_f)$, which reads as
$$
  y = \frac{-p^*\Delta^3}{C_1} ( I(\rho)+C_2),
$$
where the integration constants can be found through boundary conditions: $C_2 = -I(\rho_i)$ and $C_1=-p^*\Delta^3\left(I(\rho_o)-I(\rho_i)\right)/L$. The final approximate solution, which determines the average fluid pressure distribution along the channel coordinate $y$ is given below:
\be
  \frac y L = \frac{I(\rho) - I(\rho_i)}{I(\rho_o)-I(\rho_i)}.
  \label{eq:py}
\ee
Resulting curves for the variation of hydrostatic pressure, mean gap and contact half-width along the channel are depicted in Fig.~\ref{fig:3}, \ref{fig:4}(a), \ref{fig:4}(b), respectively.
This approximate result is able to capture the non-linear hydrostatic pressure distribution along the channel, to account for the induced deformation of the solid and thus to obtain the narrowing of the channel.
Note that the obtained fluid pressure distribution becomes a concave function and resembles the pressure distribution of a compressible fluid.

With the derived analytical solution we may calculate the fluid flux in the y-direction as
\begin{equation}
\label{eq:flux_approx}
q_y(x, y) = -\frac{g^3(x, y)}{12\mu} \frac{dp(y)}{dy},
\end{equation}
where the gap $g(x, y)$ is obtained in each section $y=\text{const}$ using the Westergaard's solution~\eqref{eq:west_disp} corresponding to the mean pressure $\bar{p} = \pext - p_f(y)$, and the derivative $dp/dy$ is calculated using~\eqref{eq:py} as
\begin{equation}
 \label{eq:dp_dy}
 \frac{dp}{dy} = \left(-\frac{1}{p^*}\frac{L}{I(\rho_o) - I(\rho_i)} \frac{dI(\rho)}{d\rho} \right)^{-1}.
\end{equation}
This result will be used in the following for computation of the mean flux and the effective transmissivity of the interface and comparison with the numerical solution.

\section{Numerical coupling schemes for fluid-solid interaction\label{sec:coupling}}

Two different approaches for numerical solution of fluid-structure interaction problems (including the fluid transport through the contact interface) are generally distinguished: the first is the \textit{partitioned} approach, under which solvers for solid mechanics problem and for the fluid transport are separated and work sequentially, thus the data exchange between them must be established; the second is the \textit{monolithic} approach, which operates with a single combined solver for both physics, so that solutions of the solid and fluid problems are obtained simultaneously. 

In this study we use the monolithic approach, some details of its implementation in the finite element framework are presented below. To satisfy the contact constraints, we add contact elements on the potential contact zone, here we utilize the mortar approach and the augmented Lagrangian method to fulfill the constraints~\cite{AlartCurnier,yastrebov2013b,wriggers2006,puso2004mortar}. To solve the Reynolds problem, fluid pressure values are added as degrees of freedom to the surface nodes of the structural mesh, and the surface elements for solving the fluid problem are appended to this mesh. Finally, to enforce the Neumann boundary conditions of the fluid pressure in the out-of contact zones of the surface, we use the mortar method again~\cite{yang2009mortar}. 
Therefore, the equations governing contact constraints, fluid flow and fluid/structure interaction are embedded into a single global system of equations, which is solved at each iteration of the Newton-Raphson method, required due to intrinsic nonlinearity of the coupled problem. 
When convergence criteria are satisfied for displacement and pressure degrees of freedom, the solution of the coupled problem is obtained for given boundary conditions. 

Note that contrary to many multiphysical coupled problems, in which different physics require significantly different temporal and spatial discretizations, the particular coupling considered here deals with stationary equations for the fluid and the solid, which can be resolved numerically on almost coinciding surface meshes. The mesh used for the fluid is obtained by projecting the surface of the solid mesh on the rigid flat, with which the solid comes in contact.

\begin{figure}
  \includegraphics[width=1\textwidth]{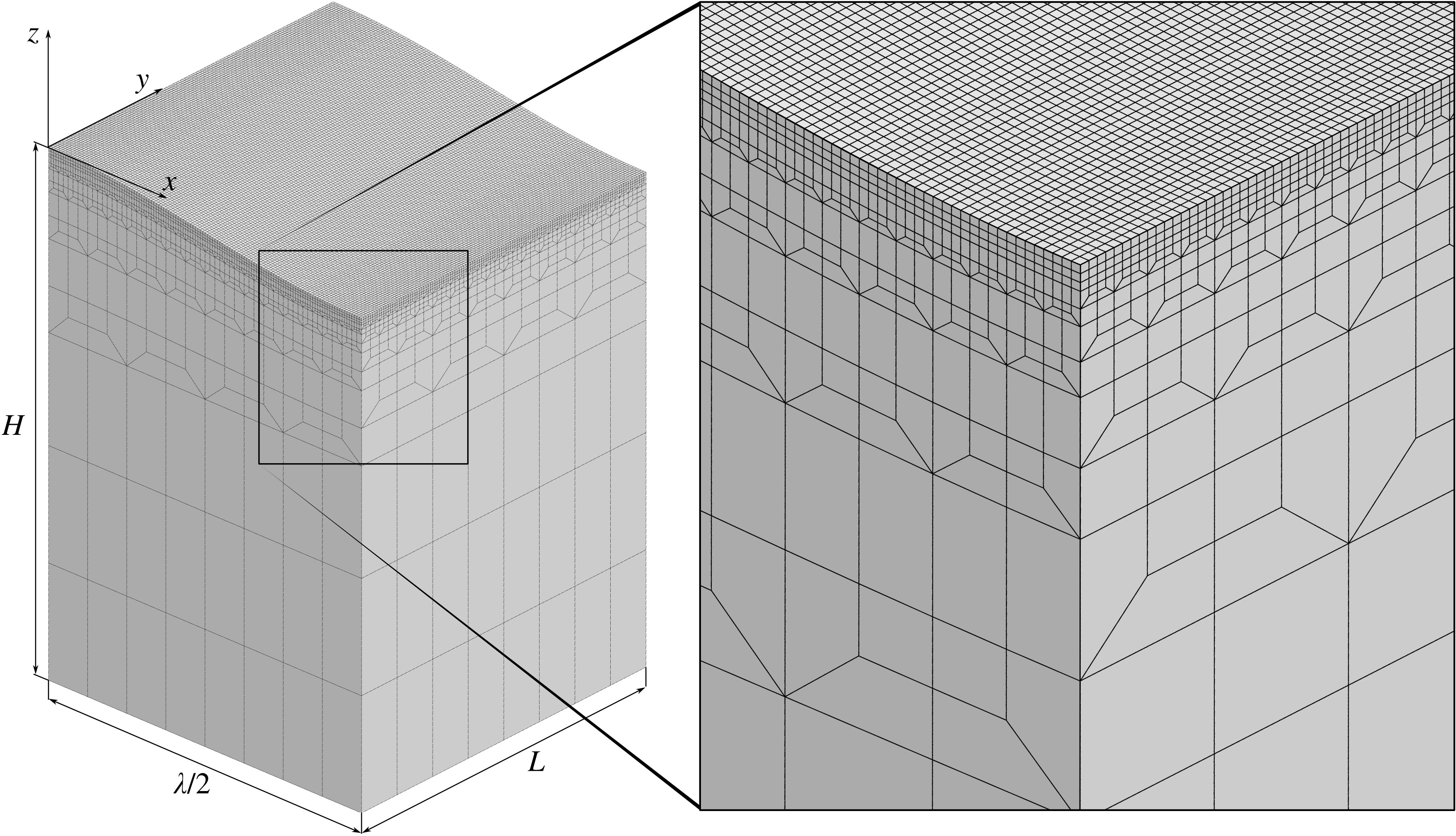}
 \caption{\label{fig:mesh}Finite element mesh with $128\times128$ elements on the contact surface, which was used to solve the coupled solid-fluid problem.}
\end{figure}

The coupling algorithm was implemented in the in-house finite element software Z-set~\cite{Besson1997,zset}.
Due to the reflection symmetry of the geometry and loads, only the half wavelength is simulated using structured finite element mesh of hexahedral linear elements, with 128$\times$128 square-shaped faces on the surface, the mesh gradually coarsens with the depth. The mesh is depicted in Fig.~\ref{fig:mesh}, it contains approximately 109\,000 nodes and approximately 98\,000 elements with 8 integration points per element. At the bottom surface $z=-H$, the displacement vector is prescribed as $u_x=u_y=0, u_z = kt$, where $k$ is a load factor, $t$ is the time. Geometrical parameters of the problem are the following $L=1$ cm, $\lambda=2$ cm, $\Delta=0.2$ mm, $H=1.4$ cm, Young's modulus $E=2$ GPa and Poisson ratio $\nu=0.3$, thus $E^* \approx 2.2$ GPa.

\section{Results\label{sec:results}}

\begin{figure}[t]
	\centering
	\includegraphics[width=0.99\textwidth]{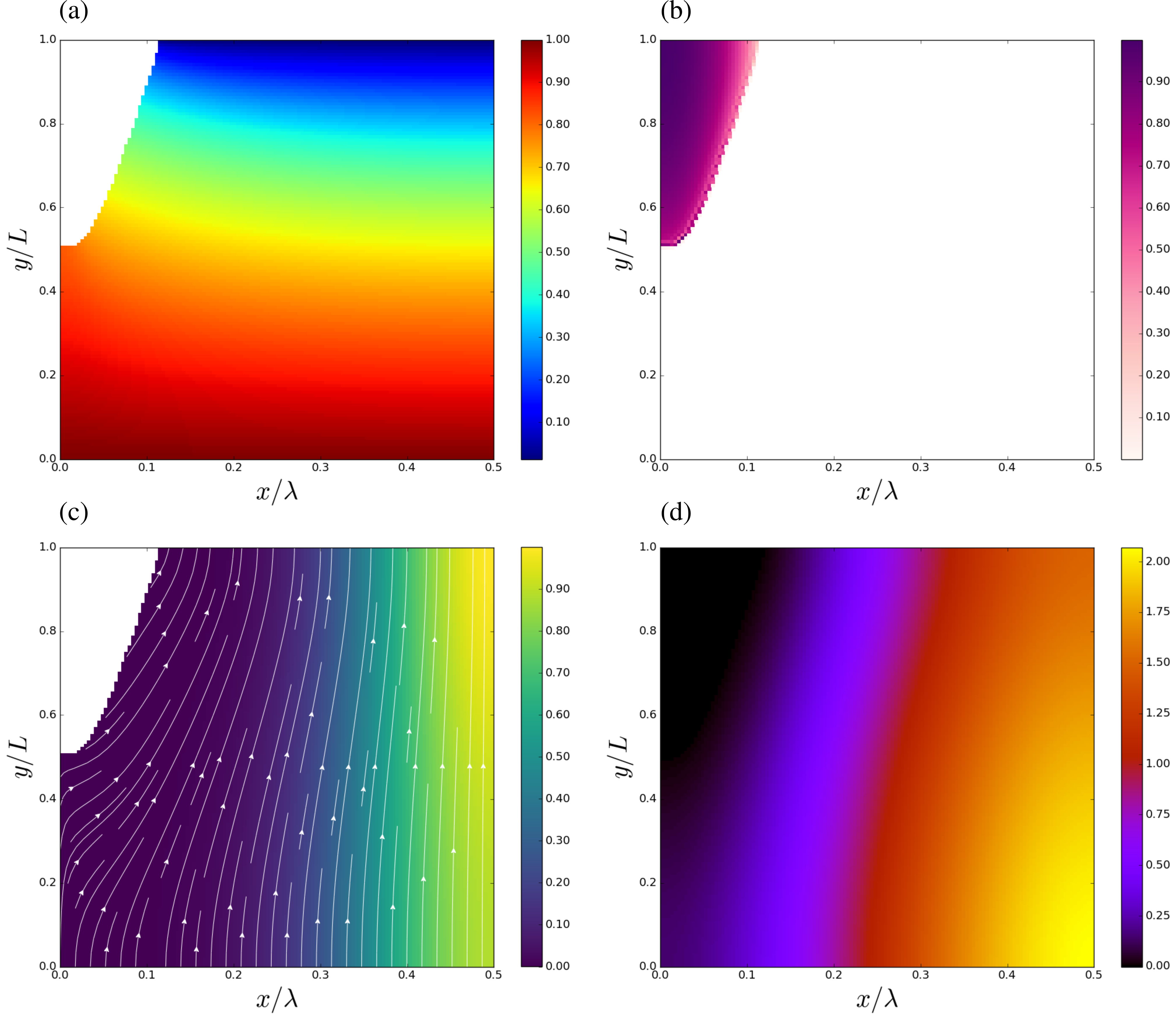} 
	\caption{Results of the numerical simulation: (a) normalized fluid pressure $p_f/p_i$, (b) normalized contact pressure $|\sigma_n|/|\sigma_n|^{\small{max}}$, (c) normalized fluid flux intensity $|q|/|q_{max}|$, (d) normalized gap $g/\Delta$; $p_{\mbox{\tiny ext}}/p^* = 0.48$, $p_{i}/p^* = 0.72$, which corresponds to $p_i/p_{\mbox{\tiny ext}} = 1.5$.}
		\label{fig:fea_result_1}
\end{figure}

\begin{figure}[t]
	\centering
	\includegraphics[width=0.99\textwidth]{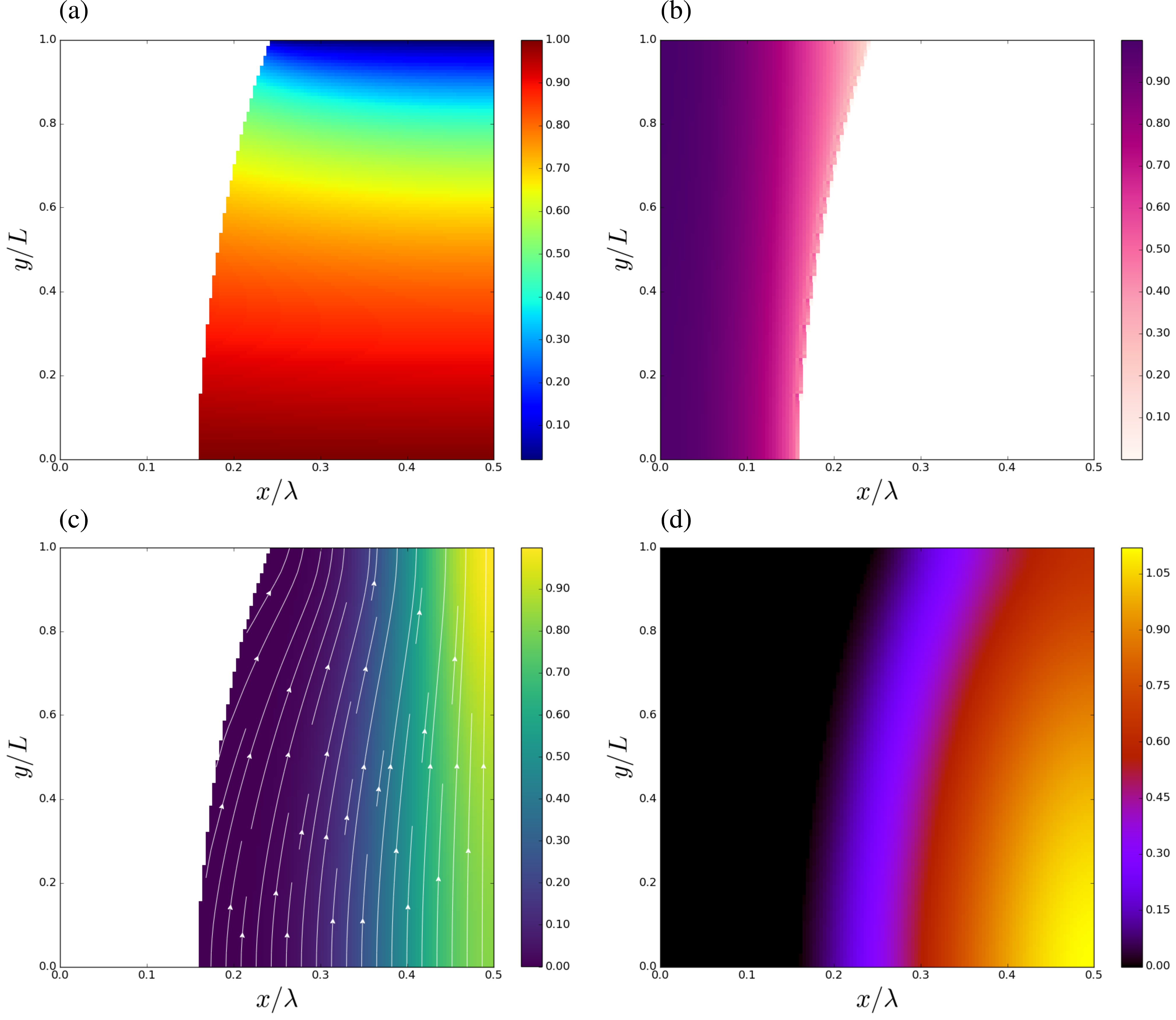} 
	\caption{Results of the numerical simulation: (a) normalized fluid pressure $p_f/p_i$, (b) normalized contact pressure $|\sigma_n|/|\sigma_n|^{\small{max}}$, (c) normalized fluid flux intensity $|q|/|q_{max}|$, (d) normalized gap $g/\Delta$; $p_{\mbox{\tiny ext}}/p^* = 0.8$, $p_{i}/p^* = 0.72$, which corresponds to $p_i/p_{\mbox{\tiny ext}} = 0.9$.}
	\label{fig:fea_result_2}
\end{figure}

We carried out several simulations for three different fixed inlet fluid pressures $p_i = 2, 10, 50$ MPa, respectively, and for outlet pressures $p_o/p_i = \{0;\, 0.25;\,\\ 0.5;\, 0.75\}$. The solid is gradually brought in contact by applying vertical displacement on its bottom side. We present the detailed results of numerical simulations in Fig.~\ref{fig:fea_result_1} and Fig.~\ref{fig:fea_result_2}: the distribution of the fluid pressure and the contact pressure, as well as the gap and the fluid flux for the case $p_i = 50$~MPa, $p_o = 0$ and for two particular load steps: $p_{\mbox{\tiny ext}}/p^* = 0.48, 0.8$, respectively. On the initial stage of loading no contact occurs and the load is supported by the fluid solely, which flows along the entire channel (this classical situation is not presented in the figures). 
For higher loads, the solid comes in contact with a rigid flat, the contact starts from the outlet zone (see Fig.~\ref{fig:fea_result_1}). 
With the increasing load the contact zone spreads out and at a certain load reaches the inlet zone (see Fig.~\ref{fig:fea_result_2}); starting from this moment the approximate solution~\eqref{eq:py} becomes applicable. The gap, being bigger at the inlet due to higher inlet pressure, narrows towards the outlet region. The flow is localized within the trough of the wavy profile, and the flux intensifies towards the outlet. Due to the narrowing of the channel along the flow direction, the current lines converge towards the outlet, it results in a small but still distinguishable fluid pressure gradient in the $OX$ direction, which was not explicitly taken into account in the derivation of the approximate solution. 

\begin{figure}[h]
	\centering
	\centering
	\includegraphics[width=0.99\textwidth]{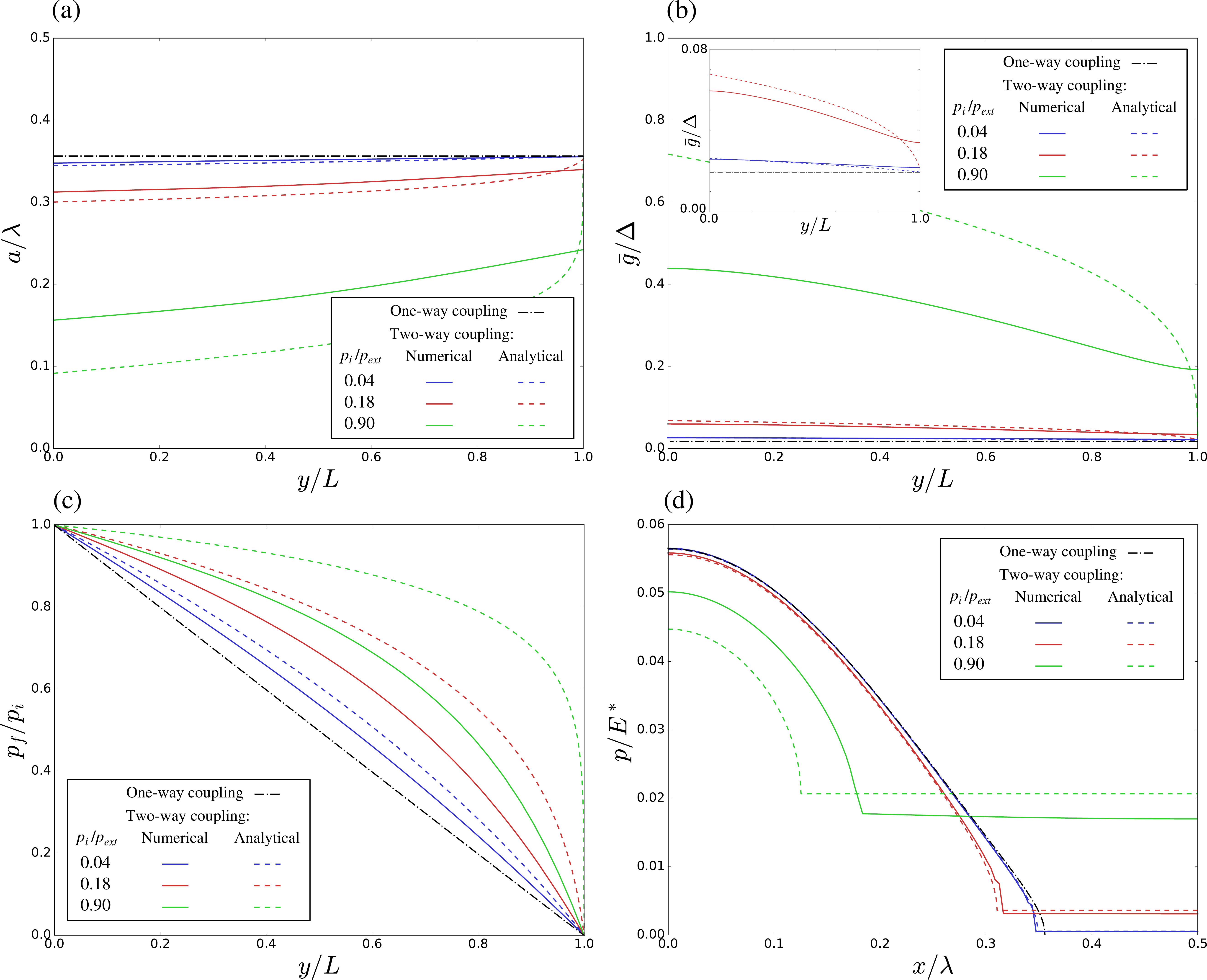} 
	\caption{Comparison of results of numerical simulations with the approximate analytical solutions for (a) normalized contact region half-width $a/\lambda$, (b) normalized mean gap $\bar g/\Delta$ and (c)  normalized fluid pressure $p_f/p_i$ evolution along the channel, (d) normalized contact and fluid pressure profiles $p/E^*$ in the section $y=L/2$:  $p=|\sigma_n|$ in $\Omega_c$, $p=p_f$ in $\Omega_f$ and $p=|\sigma_n|=p_f$ at $\partial \Omega_c\cap\partial\Omega_f$.}
	\label{fig:fea_results}
\end{figure}

In Fig.~\ref{fig:fea_results} numerical results for the variation along the channel of mean fluid pressure, mean gap and the contact width, as well as the contact and fluid pressure profiles in the section $y=L/2$ are compared with approximate solution from Section~\ref{sec:approximate}. A rather good agreement is obtained in the range of validity of the approximate result: the inlet pressure and external pressure satisfy the following conditions $p_{i} \leq p_{\mbox{\tiny ext}}$, $p_{\mbox{\tiny ext}} < p^*$, and they are chosen such that the contact zone reaches the inlet, see~\eqref{eq:pres_cond} and the discussion in Section~\ref{sec:approximate}. These limitations are quite strong and in reality ensure only a limited range of validity of the approximate solution. 

\begin{figure}[p]
	\centering
	\includegraphics[width=0.85\textwidth]{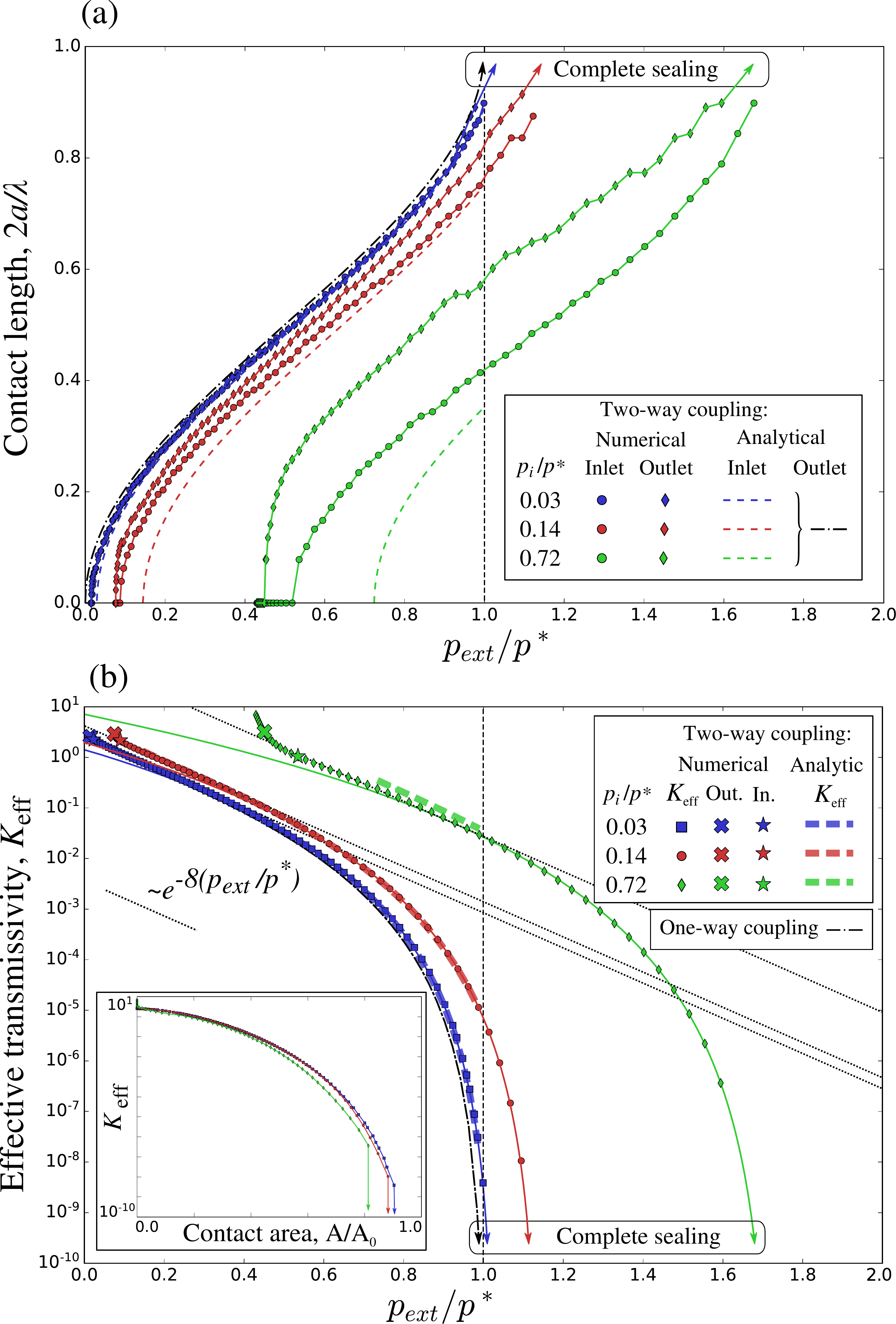}
	\caption{\label{fig:resnew} The evolution with the increasing external pressure of: (a) the length of the contact zone on the inlet and the outlet sides; (b) the effective transmissivity $K_{\text{eff}}$ of the interface for three different values of the inlet fluid pressure $p_i/p^*$ and zero outlet pressure $p_o =0$. Analytical results~\eqref{eq:flux_approx}-\eqref{eq:dp_dy} are shown as thick dashed curves, numerical results are presented using markers, while full curves are fittings of power law $K_{\text{eff}} \sim (p_c - p_{\text{ext}})^\gamma$, where $\gamma = 6\pm0.5$. ``Cross'' and ``star'' markers are used to highlight the external pressure necessary for the contact to appear at the outlet and inlet sides, respectively. Inset in (b) shows the evolution of the effective transmissivity with respect to the ratio of the real contact area to the apparent one.}
\end{figure}

In Fig.~\ref{fig:resnew}(a) the numerically computed evolution of the length of the contact zone on the inlet and the outlet sides with the increasing pressure is depicted for three different inlet pressures $p_i/p^* = 0.03, 0.14, 0.72$ and $p_o = 0$. These results are compared with the analytical solution, which is valid if the external pressure is in the interval $p_i < p_{ext} < p^*$. In accordance with the assumptions of the approximate solution, the evolution of the length of the contact zone on the outlet side is independent of the inlet fluid pressure. However, in the numerical results curves both for inlet and outlet sides shift into the region of higher external pressures with the increasing inlet pressure. Note also that in the numerical results the outlet contact length grows faster than the inlet one. Finally, the strongly coupled numerical model shows that the higher is the inlet pressure $p_i$, the higher external load is needed to completely seal the channel.

In Fig.~\ref{fig:resnew}(b) the effective transmissivity $K_{\text{eff}}$ of the wavy channel brought in contact is plotted: it is defined as
\be
  K_{\text{eff}} = -\frac{12\mu Q L}{\Delta^3 (p_o-p_i),}
\ee
where $Q$ is the mean flux over the area $\lambda \times L$, i.e. 
\begin{equation}
\label{eq:mean_flux}
	Q=\frac{1}{\lambda L}\int\limits_{0}^\lambda\int\limits_{0}^L q_y\, dxdy.
\end{equation}
For computation of the local flux $q_y(x,y)$ we used the results of numerical simulations as well as the approximate solution~\eqref{eq:py}-\eqref{eq:dp_dy}. We considered the same three cases as before with different inlet pressures. For each case we highlight the corresponding external pressure necessary for the contact to appear at the inlet and outlet sides. As soon as the contact appears on both sides, the evolution of the transmissivity becomes exponential with respect to the external pressure normalized by $p^*$ with the exponent $\approx-8$ in all three cases. We note that this coefficient is lower than the one observed in simulations of the interface transmissivity for the surfaces with representative random roughness, where it was reported to be of order $\approx -12$~\cite{dapp2016fluid}. Closer to the complete sealing (the percolation limit) the transmissivity decays faster and can be very accurately described by a power law with respect to the difference between the critical external pressure $p_c$, necessary to seal the channel, and $p_{\text{ext}}$, i.e. $K_{\text{eff}} \sim (p_c - p_{\text{ext}})^\gamma$. Note that for our results $\gamma$ was estimated as $6\pm0.5$, while in accurate but one-way coupled studies of the percolation limit of bi-sinusoidal surfaces it was found to be equal to $3.45$~\cite{dapp2015contact,dapp2016fluid}.

The transmissivity for the lowest inlet pressure $p_i/p^* = 0.03$ almost coincides with that of the one-way coupling analysis and is well described by Kuznetsov's analytical solution. For higher inlet pressures we obtain significantly bigger transmissivity. In the region of the exponential decay in case of two-way coupling we have $\approx 1.6$ times higher transmissivity than in the case of the one-way coupling for $p_i/p^* = 0.14$ and $\approx 32$ times higher for $p_i/p^* = 0.72$. 
Note that the transmissivity curves based on the analytical approximation~\eqref{eq:flux_approx}-\eqref{eq:dp_dy} are in a very good agreement with the numerical ones in the range of the validity of the former. However, the analytical result cannot be used to study the flow near the percolation since the pressure needed to seal the channel given by the analytical solution (i.e. $p_{\text{ext}}=p_*+p_o$) strongly underestimates the real one, which can be accurately studied using the numerical approach.

In the inset of Fig.~\ref{fig:resnew}(b) we plot the effective transmissivity with respect to the real contact area fraction, curves coincide for three different cases in the beginning of loading, while the complete sealing occurs at different values of the real area fraction $A/A_0 = 80\text{\%}-90\text{\%}$. Note that in one-way coupled studies of the percolation limit of the \textit{randomly rough} surfaces in contact, the complete sealing was found to correspond to $\approx 42$\% of the real contact area~\cite{dapp2012prl}.

\begin{figure}[p]
	\centering
	\centering
	\includegraphics[width=0.7\textwidth]{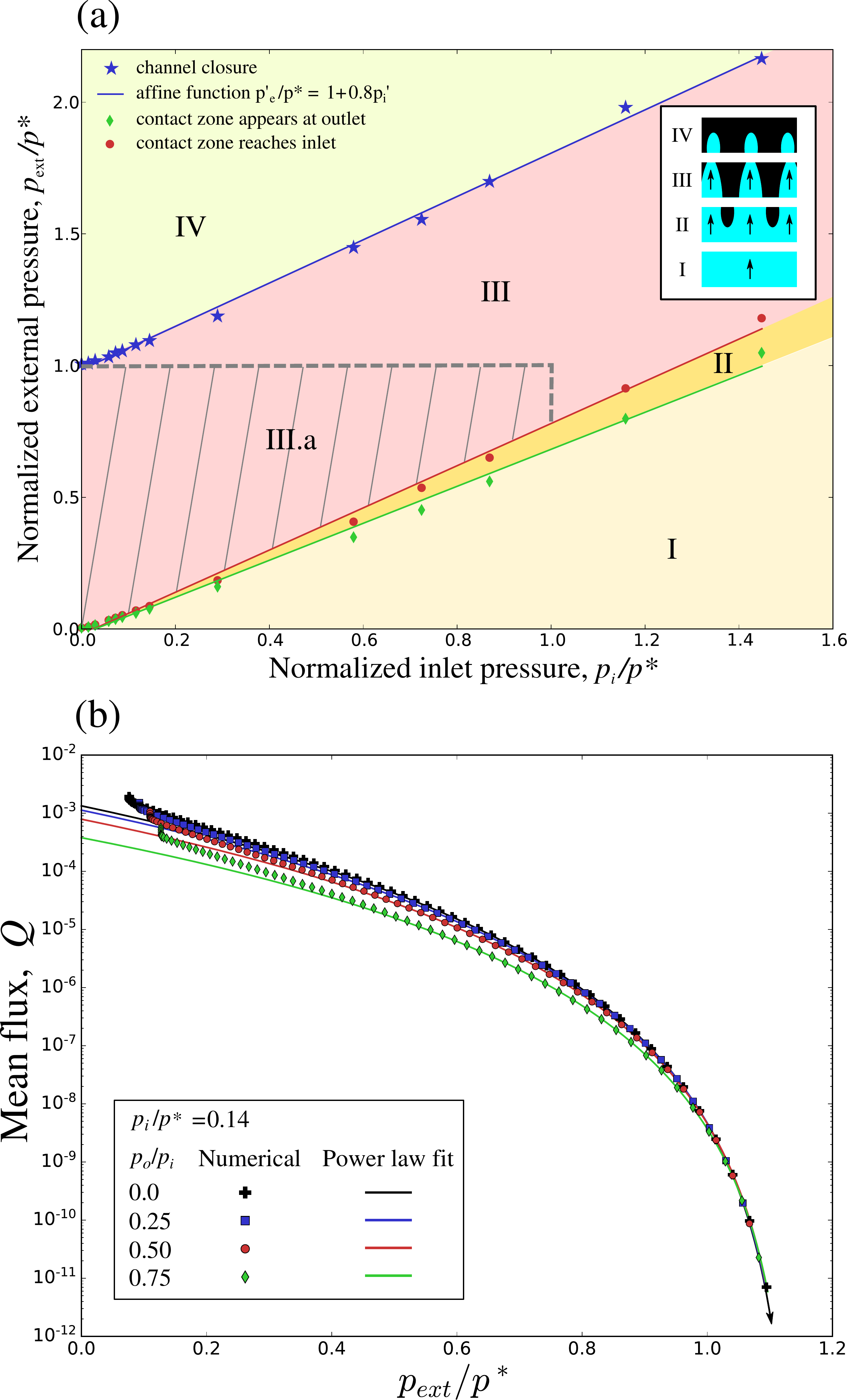} 
	\caption{(a) The phase diagram found for $p_o = 0$ demonstrating different flow regimes: I - no contact occurs in the interface, the solids are separated by the fluid flow, the external pressure is balanced by the fluid pressure; II - the contact occurs near the outlet but does not yet reach the inlet, thus the initially continuous flow fingers toward the outlet; III - the contact zones connect the inlet and the outlet forming separate channels for the fluid flow; zone III.a corresponds to the combination of loads for which the approximate analytical solution remains valid; IV - the contact is completely sealed, no flow passes through, the fluid is under the inlet hydrostatic pressure and ensures some load-bearing capacity in the non-contact region. (b) Evolution of the mean flux across the interface with the increasing external pressure for $p_i/p^* = 0.14$ and four cases of the outlet pressure $p_o/p_i = 0, 0.25, 0.5, 0.75$, showing that the critical pressure, necessary to seal the channel, is independent of the outlet pressure. Markers represent numerical results, while full curves are fittings to the power law discussed above.}
	\label{fig:fea_results_map}
\end{figure}

Finally, we evaluated the critical external pressure necessary to seal the channel, i.e. to prevent the fluid flow across the interface. The results are presented in Fig.~\ref{fig:fea_results_map}(a). The relationship between the inlet pressure and the critical external pressure is found to be close to linear, the results of the least squares fitting are presented in Fig. ~\ref{fig:fea_results_map}(a). So the critical external pressure $p_c$ needed to seal the channel can be approximately found as:
\be
p_c \approx p^* + 0.8p_i,
\label{eq:closure}
\ee
where $0.8$ is a fit parameter. We performed additional simulations with different outlet pressures for a given inlet one. The results are presented in Fig.~\ref{fig:fea_results_map}(b), where we plot  evolution of the mean flux $Q$~\eqref{eq:mean_flux} with the increasing external pressure. In the beginning of loading the mean flux is lower in case of a smaller pressure drop, however, under the increasing external load curves converge, and complete sealing of the channel occurs at the same value of the critical external pressure, which therefore is determined only by the inlet pressure. The explanation comes from the fact that close to the complete sealing of the channel, the fluid pressure drop occurs in the vicinity of the outlet, while in the remaining channel the pressure is close to the inlet one. Therefore, remarkably, the load-carrying capacity of the fluid in the interface close to percolation is defined only by the inlet pressure.

In addition, in Fig.~\ref{fig:fea_results_map}(a) we highlight the external pressure at which the contact appears on the inlet side, this corresponds to the onset of validity of the approximate solution. The end of its validity corresponds to $p_{\mbox{\tiny ext}}/p^* = 1$. The narrow validity range of our approximate solution can be improved by dropping the assumption (ii) in Section~\ref{sec:approximate} and by including strong elastic interaction between $y$-sections, which would lead to a much more complicated analysis and is not addressed here.

\section{Concluding remarks\label{sec:conclusion}}

We presented a theoretical study of the pressure driven creeping flow in contact interface formed between solids with wavy surfaces. We considered the case when the fluid flows across the wavy section in channels delimited by mechanical contact zones. This problem is relevant for certain applications of thin fluid flow in contact interfaces, including sealing, hydrogeology and biological systems. A strong (two-way) coupling between fluid flow and deformation of the solid was assumed, which is crucial for applications in which the fluid pressure is comparable with the mean contact pressure, for example, for the soft matter or biological tissue.

We derived an approximate analytical solution based on the Westergaard-Kuznetsov solution and a one-dimensional formulation of Reynolds equation. This solution describes both the solid deformation and the fluid pressure distribution in the strongly coupled case. A finite-element monolithically coupled framework for fluid and solid equations was also used to solve this non-linear multi-field problem and to prove the validity of the approximate solution. Despite a rather limited interval of loads within which the latter is applicable, it can provide a useful first-order approximation for the analysis of transmissivity of contact interfaces. At the same time, numerical results showed that in a wide range of the external loads up to the complete sealing of the channel, the transmissivity of the interface can be described by a power law, which has already been reported in the studies of contact interfaces having representative and model roughness.

Both numerical and analytical results, which take into account two-way coupling, showed that the interface transmissivity is significantly higher than this predicted by the one-way coupling if the fluid pressure is high enough. An additional result of this study is the affine dependence of the external critical pressure which seals the channel on the inlet fluid pressure: this relation may be shown useful in sealing applications as well as in soft porous or cracked media, in zones where the flow can be described by Reynolds equation. Remarkably, this critical pressure was found to be independent of the outlet pressure.

\section*{Acknowledgments}
The authors are grateful to Julien Vignollet for useful comments to the paper and to Andreas Almqvist and Francesc P\'{e}rez R\`{a}fols for enriching discussions. The financial support of Safran Group and MINES ParisTech (Th\`{e}se-Open) is acknowledged.


\end{document}